\documentclass[twocolumn]{webofc}
\usepackage[varg]{txfonts}   
\usepackage[hidelinks]{hyperref}
\usepackage{cleveref}
\usepackage{booktabs}
\usepackage[font=small,labelfont=bf]{caption}
\let\clearpage\relax
\begin{document}
\title{Investigating the effect of porosity on the soil water retention curve using the multiphase Lattice Boltzmann Method}
%
%

\author{\firstname{Reihaneh} \lastname{Hosseini}\inst{1}\thanks{\email{reihos@utexas.edu}} \and
        \firstname{Krishna} \lastname{Kumar}\inst{1}\thanks{\email{krishnak@utexas.edu}} \and
        \firstname{Jean-Yves} \lastname{Delenne}\inst{2}
}

\institute{The University of Texas at Austin, 301 E Dean Keeton St, Austin, TX 78712
\and
           IATE, Univ. Montpellier, CIRAD, INRAE, Montpellier SupAgro, Montpellier, France
          }

\abstract{%
The soil water retention curve (SWRC) is the most commonly used relationship in the study of unsaturated soil. In this paper, the effect of porosity on the SWRC is investigated by numerically modeling unsaturated soil using the Shan-Chen multiphase Lattice Boltzmann Method. The shape of simulated SWRCs are compared against that predicted by the van Genuchten model, demonstrating a good fit except at low degrees of saturation. The simulated SWRCs show an increase in the air-entry value as porosity decreases.
}
\maketitle
\section{Introduction}
\label{intro}
The  Soil Water Retention Curve (SWRC) correlates the degree of saturation or water content of unsaturated soil to its matric suction or capillary pressure. Idealized functional forms of SWRC~\cite{Brooks1964a,VanGenuchten1980a} are often used to estimate mechanical properties of unsaturated soil such as effective stress and shear strength~\cite{Khalili1998a,Lu2006a} as well as hydraulic properties such as relative permeability~\cite{Mualem1976,FREDLUND1994a}. As the soil undergoes volumetric deformation during shearing, its water retention behavior changes with the change of porosity. Many researchers have proposed models that consider the effect of porosity (or void ratio) on SWRC~\cite{Aubertin1998a,Gallipoli2003a,Tarantino2009a,Masin2009a,Zhou2012a}. In this study, we investigate the effect of porosity on the suction characteristics of unsaturated soil from a micromechanics perspective~\cite{Richefeu2016a}, by simulating the multiphase system using the Shan-Chen Lattice Boltzmann Method (LBM).

\section{Numerical method}
\begin{figure}[htbp]
    \centering
    \includegraphics[width=0.9\columnwidth,clip]{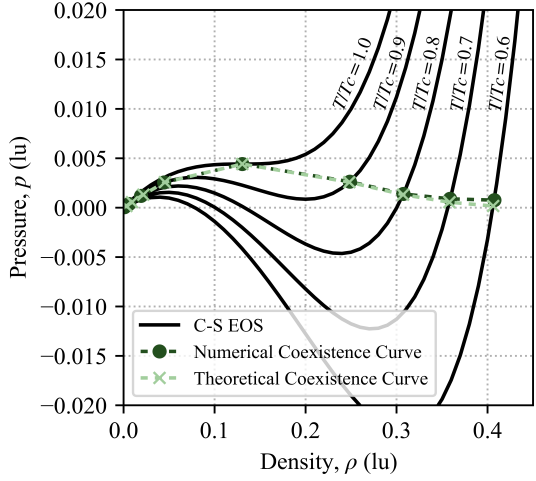}
    \caption{Comparison of the numerical (multiphase LBM simulation) and theoretical coexistence curves for the C-S EOS at different $T/T_c$.}
    \label{coex_curve}
\end{figure}

An in-house 3D multiphase LBM code is developed for modelling the multiphase fluid domain of unsaturated granular material. The D3Q19 scheme~\cite{Qian1992} for velocity discretization and the BGK collision operator are used. The interaction between the three phases, i.e. liquid, gas and solids, is considered by introducing Shan-Chen (SC)-type interaction forces into the model \cite{Shan1993a,Shan1994}. The SC interaction forces per unit volume are calculated using
\begin{equation}\label{eq_1}
\boldsymbol{F}_{SC}(\boldsymbol{x})=-\psi(\boldsymbol{x})G\sum_iw_i\psi(\boldsymbol{x}+\boldsymbol{c_i}\Delta t)\boldsymbol{c_i}\Delta t,
\end{equation}
where $\psi$ is an effective density, $G$ is a parameter that controls the strength of the interaction (negative for attraction), $\{\boldsymbol{c_i}\}$ is the discrete velocity set with $\{w_i\}$ as the corresponding weights, and $\Delta t$ is the time step which is set to 1 dimensionless lattice unit (lu). $\psi$ is defined as
\begin{equation}\label{eq_2}
\psi(\rho)=\sqrt{\frac{2}{c_s^2\Delta t^2G}(p(\rho)-\rho c_s^2)} ,  
\end{equation}
where $c_s$ is lattice sound speed equal to $\frac{1}{\sqrt{3}} \frac{\Delta x}{\Delta t}$ for the D3Q19 model~\cite{Qian1992,He1997} with both $\Delta x$ and $\Delta t$ set to 1 lu, $\rho$ is density at point $\textit{\textbf{x}}$, and $p$ is pressure. ~\Cref{eq_2} is a rearrangement of the Equation of State (EOS) of the SC-type multiphase system, which allows incorporating different EOS by simply redefining $p$~\cite{Yuan2006a}. The Carnahan-Starling (C-S) EOS is used to increase the numerical stability for systems with large liquid-gas density ratios:
\begin{equation}\label{eq_3}
p=\rho RT\frac{1+b\rho/4+(b\rho/4)^2-(b\rho/4)^3}{(1-b\rho/4)^3}-a\rho^2,
\end{equation}
with $a = 0.4963R^2T_c^2/p_c$ and  $b = 0.18727RT_c/p_c$~\cite{Yuan2006a}. $T_c$ is the temperature below which phase separation occurs and $p_c$ is the pressure at which the first and second derivatives of the EOS at $T_c$ are zero. The C-S EOS curves for different temperatures are shown in~\Cref{coex_curve}.

The developed code is validated by comparing the coexistence densities, the densities at which the liquid and gas phases coexist in a multiphase system, from the numerical simulations against the theoretical solution, at the different $T/T_c$ values shown in~\Cref{coex_curve}. The density at each node is initialized to a value between the liquid and gas densities, with a small random perturbation to allow phase separation, and the system is allowed to equilibrate. Once a liquid droplet is formed, the average liquid and gas densities are measured, and the corresponding pressures are calculated using~\Cref{eq_3}. For this study $a = 1$ lu, $b = 4$ lu and $R = 1$ lu are used~\cite{Yuan2006a}.~\Cref{coex_curve} shows a good match between the numerical and the theoretical solutions.

\section{Model description and simulation}
In this study, we use four different grain configurations with initial porosities ($\eta=V_{void}/V_{total}$) of 0.50, 0.40, 0.30, 0.15 (corresponding to void ratios of 1.00, 0.67, 0.43, 0.18) to investigate the effect of porosity on the SWRC. These configurations are created by randomly positioning grains inside a fixed-size domain until the required porosity is achieved. As the objective of this study is to develop SWRCs at constant porosities, the grain positions are fixed and grain-grain interactions are not considered. The granular assembly comprises of grain diameters between 10 lu and 30 lu, with a uniform distribution. The lattice spacing, $\Delta x$, is set to 1 lu, which corresponds to a minimum of 10 grid points per grain diameter. The LBM domain is set to $200\times200\times200$ lu$^3$, with periodic boundaries in every direction to eliminate boundary effects. 

In all simulations, the relaxation parameter, $\tau$, for the BGK collision operator is set to 1 lu. The parameters for the C-S EOS are chosen as $a = 1$ lu, $b = 4$ lu and $R = 1$ lu \cite{Yuan2006a}, corresponding to $T_c = 0.0943$ lu, $p_c = 0.0044$ lu, and $\rho_c = 0.13044$ lu. $T/T_c$ of 0.7 is selected for these simulations. $G$ is set to -1 (by using~\Cref{eq_2} for $\psi$, the parameter $G$ cancels out when this equation is substituted in~\Cref{eq_1}, therefore, the magnitude of $G$ is irrelevant and only its sign is of importance). The density of solids, $\rho_s$, in a multiphase simulation only controls the contact angle between the solids and the liquid and does not correspond to the physical density of the solids; $\rho_s$ closer to the liquid density, $\rho_l$, results in a more hydrophilic surface, whereas  $\rho_s$ closer to the gas density, $\rho_g$, creates a more hydrophobic surface \cite{Kruger2017}. In these simulations, $\rho_s$ is set to 0.98$\rho_l$, corresponding to a contact angle of 14$^{\circ}$. 

The dimensionless lattice units, lu,  used for all the parameters in this study can be converted to physical units by choosing appropriate conversion factors. It can be shown that the fluid kinematic viscosity in an LBM simulation is $\nu=c_s^2(\tau-0.5)$ in lu~\cite{Qian1992}; the conversion factors for length, $C_l$, and time, $C_t$, should be chosen such that $c_s^2(\tau-0.5)C_l^2/C_t$ corresponds to the correct physical kinematic viscosity. In addition to this criteria, accuracy and stability should be taken into account when choosing these conversion factors. For this qualitative study, all values are simply reported in lu.

In all four configurations of the granular assembly, the density of the fluid domain is initialized to a value between the liquid and gas coexistence densities (0.358 lu and 0.009 lu, respectively), with a random perturbation of $\pm0.1$, to allow phase separation. After an adequate number of steps, when the fluid has phase separated and reached equilibrium, liquid is injected into the system by increasing the density of all fluid nodes by $0.005$ lu every 1000 steps. It is ensured that 1000 steps is enough for the system to reach equilibrium before the next injection. The SWRC is generated by calculating the average suction, $\delta p$, and degrees of saturation, $\mathit{S_r}$, at the end of each injection stage. $\delta p$ is defined as the difference between the average gas pressure and the average liquid pressure in the entire fluid domain. $\mathit{S_r}$ is measured by dividing the total number of liquid nodes by the total number of fluid nodes (voids). In addition, the number of liquid clusters at each injection stage is computed using the Depth-First Search algorithm~\cite{tarjan1972depth}.

\section{Results and discussion}

\begin{figure}[htb]
\centering
\includegraphics[width=0.7\columnwidth]{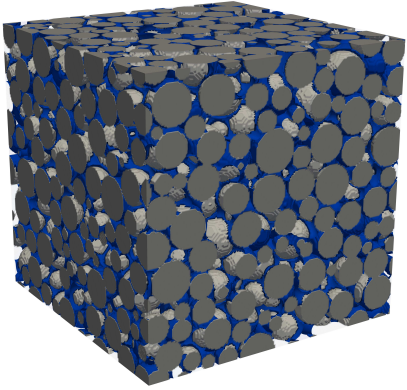}
\caption{3D view of the multiphase domain with $\eta$ = 0.40 at $\mathit{S_r}$ = 0.4.}
\label{3d}   
\end{figure}

\begin{figure}
\centering
\includegraphics[width=0.7\columnwidth]{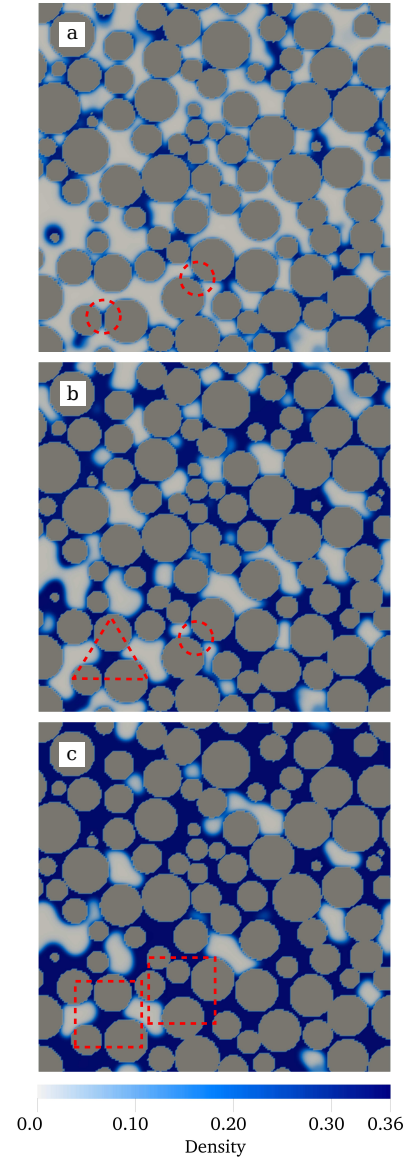}
\caption{Snapshots of 2D slices of the multiphase domain with $\eta$ = 0.40 at (a) $\mathit{S_r}$ = 0.1, (b) $\mathit{S_r}$ = 0.4 and (c) $\mathit{S_r}$ = 0.7. The dashed-circle zones show examples of binary liquid bridges, the dashed-triangle zone shows an example of a liquid cluster connecting multiple grains, and the dashed-square zones show examples of concave menisci in the capillary state.}
\label{states}
\end{figure}

\Cref{3d} shows a 3D view of the grain configuration with $\eta$ = 0.40 and the liquid clusters formed at $\mathit{S_r}$ = 0.4.~\Cref{states} shows snapshots of 2D slices of the unsaturated granular media, indicating that multiphase LBM is able to capture the different states of liquid clustering: pendular (\ref{states}a), funicular (\ref{states}b), capillary  (\ref{states}c), and droplet states~\cite{mitarai2006wet}. In the pendular state, pairs of grains are connected by binary liquid bridges (e.g. dashed-circle zones). In the funicular state, a combination of binary liquid bridges (e.g. dashed-circle zone)  and liquid clusters connecting multiple grains (e.g. dashed-triangle zone) is present. In the capillary state, the liquid almost fills the entire pore space between grains, however, its surface forms concave menisci (e.g. dashed-square zones) exerting a net cohesive force on the grains. In the droplet state, the liquid immerses the entire domain and exerts no capillary forces on the grains.

\begin{figure}[tb]
\centering
\includegraphics[width=0.9\columnwidth]{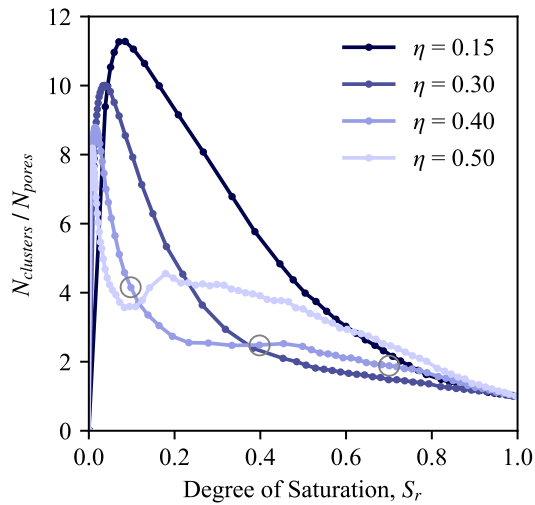}
\caption{Evolution of normalized number of clusters as a function of degree of saturation, for different porosities. The circles indicate points at which snapshots of the simulation are provided in~\Cref{states}.}
\label{cluster}
\end{figure}

The pore-scale evolution of clusters with increase in saturation is investigate by examining the evolution of number of liquid clusters.~\Cref{cluster} shows the number of liquid clusters normalized by the number of disconnected pores in the soil structure ($N_{cluster}/N_{pores}$) as a function of $\mathit{S_r}$, for different porosities. The circles indicate points at which snapshots of the simulation are provided in~\Cref{states}. As $\mathit{S_r}$ increases from zero, $N_{cluster}/N_{pores}$ initially increases rapidly up to a peak value, due to the formation of a large number of small liquid droplets on the surface of the grains. With further increase in saturation, these small droplets aggregate together to form liquid films around the grains and subsequently coalesce to binary bridges between grains, thereby decreasing the number of individual clusters at a steep rate. The number of distinct clusters gradually decreases as the system moves from the pendular state to the funicular state, and finally, to the capillary state. Eventually, all the pores are filled with liquid, and the cluster count becomes equal to the number of disconnected pore spaces. 

As the porosity decreases, the peak of the curves in~\Cref{cluster} shift to the right, indicating that a higher $\mathit{S_r}$ is required for the initial clusters to connect and form liquid films on the grain surfaces. This is expected as lower porosity corresponds to a larger number of tiny pore spaces and a higher surface area of the granular assembly. In addition, in the case of the lowest porosity, the number of distinct clusters decreases more gradually throughout the transition from the pendular state to the capillary state, compared to the other porosity cases.

\begin{figure}[t]
\centering
\includegraphics[width=0.9\columnwidth]{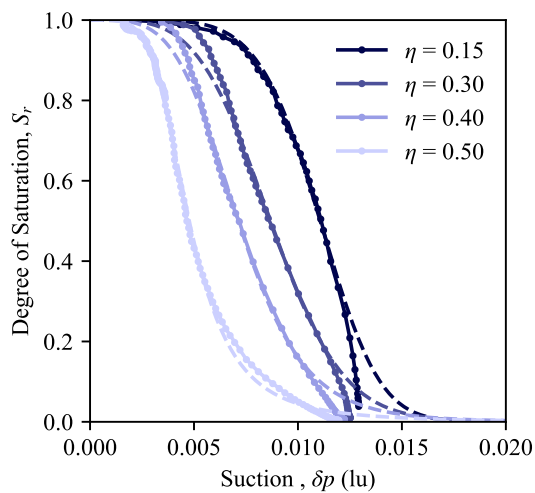}
\caption{SWRCs from the LBM simulations for different porosities. The dashed lines represent the fitted curves based on the van Genuchten model.}
\label{wrc}
\end{figure}

The SWRCs simulated for different porosities are shown in~\Cref{wrc}. The simulated SWRCs are compared against fitted curves based on the simplified van Genuchten model \cite{VanGenuchten1980a, Gallipoli2003a}
\begin{equation}\label{vG}
\mathit{S_r}=(\frac{1}{1+(\alpha\delta p)^n})^m,
\end{equation}
where $\alpha$, $n$ and $m$ are fitting parameters. The fitting parameters for the curves in this study are reported in~\Cref{tab-1}. Parameter $\alpha$ controls the shift of the curve to left and right and is inversely proportional to the air-expulsion/air-entry value (AEV). It can be deduced from the shift of the simulated curves in~\Cref{wrc} or the value of $\alpha$ in~\Cref{tab-1}, that the AEV for the simulated SWRCs increases with decreasing porosity. The inverse relation between the AEV and porosity is consistent with all proposed models for the SWRC which take void ratio into account \cite{Aubertin1998a,Gallipoli2003a,Tarantino2009a,Masin2009a,Zhou2012a}. Parameter $n$ controls the slope of the curve and is related to the pore size distribution, with higher $n$ corresponding to a more uniform distribution~\cite{Matlan2014a}. Comparing the $n$ parameter for the different curves shows that the curves with the highest and lowest porosities have steeper slopes at the middle section of the SWRC, compared to the curves with intermediate porosities. This could be explained by assuming that the lowest and highest porosity models have uniform small and large pores, respectively, while the intermediate porosity models have a combination of small and large pores; however, the verification of this assumption is left for future studies. Parameter $m$ controls the symmetry of the curve when $\delta p$ is plotted in log space. As seen in~\Cref{wrc}, the maximum suction of the simulated curves is constrained, creating an unsymmetrical shape for the SWRC at lower porosities. Therefore, $m$ is higher for lower porosities.
 
\begin{table}[htbp]
\centering
\caption{van Genuchten fitting parameters.}
\label{tab-1}
\begin{tabular}{llll}
\hline
$\eta$ & $\alpha$ & $n$ & $m$ \\\hline
0.15 & 67 & 6 & 4.50\\
0.30 & 79 & 4 & 3.47\\
0.40 & 119 & 4 & 1.66\\
0.50 & 228 & 5 & 0.75\\\hline
\end{tabular}
\end{table}

The SWRC for the simulation with the highest porosity ($\eta$ = 0.50) has the expected ``S'' shape and the fitted van Genuchten (vG) curve matches the trend of the data. For the two simulations with intermediate porosities, the fitted vG curves match the simulated curves perfectly in the $\mathit{S_r}$ range of 0.1 to about 0.8. However, at higher saturation ($\mathit{S_r}$ > 0.8), the vG curves smoothly transition to zero suction while the simulated curves continue with the same slope up to $\mathit{S_r}$ = 0.96, where the suction abruptly drops to zero. Furthermore, as the saturation decreases below 0.1 for these intermediary porosity cases, LBM simulations only indicate a very small increase in suction, which contrasts the dramatic increase predicted by vG curves. For the lowest porosity case ($\eta$ = 0.15), the smooth transition to zero suction at high $\mathit{S_r}$ occurs as expected, however, at $\mathit{S_r}$ below 0.2, the suction remains unchanged. Interestingly, all simulated curves converge to the same value of maximum suction, indicating that the maximum suction in these multiphase LBM simulations may be constrained by the modeling parameters. Further research is required to understand the cause of this behavior.

\section{Conclusion}

The effect of porosity on the Soil Water Retention Curve (SWRC) was studied by simulating unsaturated soil using the 3D multiphase Shan-Chen-type Lattice Boltzmann Method. Four different grain configurations with porosities ranging from 0.15 to 0.50 were used. The four different states of liquid clustering, namely pendular, funicular, capillary, and droplet states, were observed in the multiphase simulation for the different porosity cases. The simulations were able to correctly capture the increase of the air-entry value with decreasing porosity and the shift of the SWRC towards higher suction. However, the results indicated that the maximum suction is constrained at low degrees of saturation and further research is needed to explain this behavior.

\bibliography{mphase}
\end{document}